\documentclass[
 reprint,
 superscriptaddress,
 amsmath,amssymb,
 aps,
 prl
]{revtex4-2}

\usepackage{color}
\usepackage{amsmath}
\usepackage{pifont}% http://ctan.org/pkg/pifont
\usepackage{amssymb}  % More symbols
\usepackage{bbold}
\usepackage{float}
\usepackage{placeins}
\usepackage{tikz}
\usepackage{makecell}
\usepackage{subfigure}
\usepackage{scalerel}
\usepackage{pifont}   % Ding symbolshttps://www.overleaf.com/project/6084801cec8058e65885c58b
\usepackage{dcolumn}  % Align table columns on decimal point
\usepackage{multirow} % Table functions
\usepackage{placeins}% For making floats not move around everywhere
\usepackage{soul}
\usepackage{xcolor}
\usepackage{graphicx}% Include figure files
\usepackage{dcolumn}% Align table columns on decimal point
\usepackage{bm}% bold math
\usepackage{listings}
\usepackage[dvipsnames]{xcolor}

\begin{document}

\preprint{APS/123-QED}

\title {Low frequency electric field sensing with a Rydberg beam}

\author{Jeremy Glick}
\affiliation{DEVCOM Army Research Laboratory South, Austin, Texas 78712 USA
\looseness=-1}

\author{John R Dickson}
\affiliation{Department of Physics, Center for Complex Quantum Systems, The University of Texas at Austin, Austin, Texas 78712 USA  
\looseness=-1}

\author{Josie Wood}
\affiliation{Department of Physics, Center for Complex Quantum Systems, The University of Texas at Austin, Austin, Texas 78712 USA  
\looseness=-1}

\author{Paul Kunz}
\affiliation{DEVCOM Army Research Laboratory South, Austin, Texas 78712 USA
\looseness=-1}
\affiliation{Department of Physics, Center for Complex Quantum Systems, The University of Texas at Austin, Austin, Texas 78712 USA
\looseness=-1}

\date{\today}% It is always \today, today,
             % but any date may be explicitly specified

\begin{abstract}
We present a method for performing low frequency electric field sensing via ionization detection of Rydberg atoms in a collimated atomic beam. A collimated beam avoids much of the electric field screening effects that are common in warm vapor cells due to the accumulation of alkali-metal atoms on glass surfaces. Further, a beam facilitates a spatially separated region for high signal-to-noise readout via ionization detection. Using this approach, we measure DC Stark shifts from external fields with frequencies as low as 1 Hz. The sensor demonstrates a sensitivity of better than 1 mV/m$\sqrt{\rm {Hz}}$ for frequencies above 20 Hz and $0.14(4)$ mV/m$\sqrt{\rm {Hz}}$ above 500 Hz with a linear dynamic range of over 50 dB.

\end{abstract}

\maketitle

\section{Introduction}

Electrometry with Rydberg atoms has been well demonstrated over wide frequency ranges from kilohertz to terahertz \cite{Meyer_assessment_2020, Adams_KHz_2008,Raithel_RF_2016, Bonnie_RydbergSensing_2021}. Rydberg states have various properties that scale advantageously with principle quantum number n including the dipole moment $(\propto n^2)$ and polarizability $(\propto n^7)$ making them highly sensitive to electric fields. The properties of Rydberg atoms and response to electric fields are traceable to fundamental constants making them ideal for use as calibration tools and in precision measurement applications \cite{Shaffer_MicrowaveElectrometry_2013, Shaffer_RF_sensing_calibration_2015, Holloway_Calibration_2024}. Rydberg sensors can simultaneously respond to disparate frequency bands \cite{Cox_MultibandDetection_2023}, perform polarimetry \cite{Meyer_Polarimeter_2024}, image terahertz frequency fields \cite{Weatherill_THzImaging_2017, Weatherill_THzImaging_2020}, reach the standard quantum limit for individual particles \cite{ Kunz_OptimalEIT_2021,Liang_SQL_2024}, and in principle reach the Heisenberg limit when leveraging entanglement \cite{Cappellaro_QuantumSensing_2017,Facon_SensitiveElectrometer_2016}.

Current Rydberg sensors commonly use warm vapor cells containing alkali-metal atoms, which pose challenges for sensing in the extremely low frequency (ELF) band. Sensing in the ELF band has applications in geophysics \cite{Biswas_geophysics_2016, Colin_geophysics_2016}, submarine communication \cite{Ali_submarine_2020}, and instances where the detection of weak voltage signals is required. A primary challenge to ELF sensing with Rydberg vapor cells is that alkali-metal atoms can accumulate on the glass, increasing the conductivity and leading to shielding of low-frequency fields \cite{Papoyan_GlassConductivity_1999,
Adams_2007_shielding, Arimondo_RbMotShielding_2011,Yuan_LowFreqSapphire_2020}. Field plates can be placed within cells to circumvent this issue. These "port-coupled" devices require an external aperture (i.e. antenna) to collect the free-space field and couple it through a vacuum port to internal field plates \cite{Kitching_internalPlates_2022, Suotang_internalPlates_2023}. Vapor cells constructed of sapphire,  or with internal paraffin-coatings,  have shown a lower timescale of screening enabling sensitive detection at 1 kHz \cite{Kayim_lowFreq_2026} and down to single Hertz \cite{Yuan_LowFreqSapphire_2020, Yuan_ELF_2026, Chandra_ELF_2026}. 

Here, rather than a warm vapor, we create a collimated beam of Rydberg atoms inside the vacuum cell, and this offers two major benefits for ELF sensing. First, the atoms can be directed away from the glass surfaces that serve as apertures for electromagnetic fields, thus reducing issues with alkali-metal buildup and shielding effects. Second, the collimated beam readily enables high-efficiency ionization detection of the Rydberg atoms. This yields large signal-to-noise (SNR) ratio ion spectra, and high sensitivity measurements of external fields. Field ionization of Rydberg atoms has a long history, such as in studies of Rydberg-Rydberg interactions \cite{Gallagher_RydbergInteraction_1981, Gallagher_RydbergInteraction_1998, Vanhaecke_RydbergInteractions_2010}, Stark shifts of Rydberg states \cite{Tannian_StarkShift_1999, Grimmel_StarkShift_2017}, and state selective ionization of Rydberg states \cite{Kleppner_SFI_1975, Edelstein_SFI_1977, Cassidy_SFI_2018}. Rydberg ionization techniques 
have also been used for field sensing of internal charge distributions within vacuum systems \cite{Merkt_StarkShift_Cali_1999, Wallraff_Cryogenic_2015}. 

In this report we measure the DC Stark shift of Rydberg atoms in $101P_{3/2}$ to detect electric fields external to our vacuum chamber. External electric fields with frequencies down to 1 Hz are observable, and we achieve a sensitivity better than 1 mV/m$\sqrt{\rm{Hz}}$ for frequencies above 20 Hz and $0.14(4)$ mV/m$\sqrt{\rm{Hz}}$ above 500 Hz.

\section{Method}

\subsection{Design overview}

\begin{figure}
    \centering
    \includegraphics[width=0.48\textwidth]{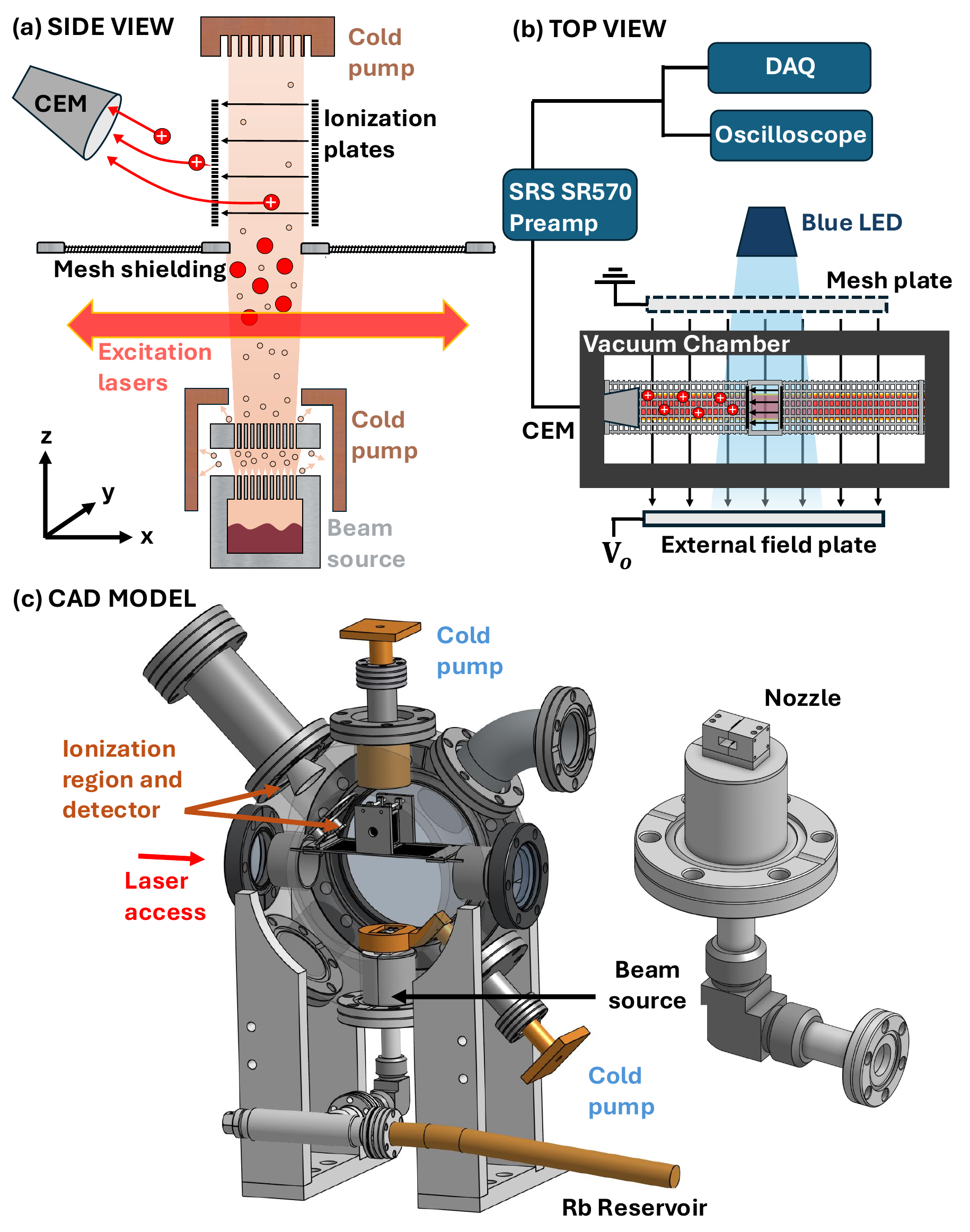}
    \caption{Overview of the experimental apparatus; (a) and (b) not drawn to scale. (a) A collimated $^{85}$Rb beam passes through lasers exciting a portion of the atoms to a Rydberg state. These atoms pass through a set of mesh plates that ionize the Rydberg atoms and deflect the ions towards a CEM detector. (b) Two plates external to the vacuum system produce a low frequency electric field at the excitation region in the chamber. A blue LED is used to reduce time varying perturbations on the ion signal. (c) CAD model of the vacuum system and atomic beam source. Some components have been made transparent to aid in visual clarity.}
    \label{fig:exp_setup}
\end{figure}

An overview of the experimental setup is shown in Fig.~\ref{fig:exp_setup}. A reservoir containing rubidium is heated, producing a vapor that flows through a nozzle and creates a collimated effusive atomic beam. This beam travels approximately 4.5 cm and passes through a laser-excitation region where a portion of the atoms are excited to a Rydberg state. The atoms then travel about 1.7 cm where they enter between two parallel mesh plates, which produce an electric field sufficient to ionize the Rydberg atoms. The positive ions then pass through the mesh and are deflected towards a channel electron multiplier (CEM) located about 6 cm from the ionization region. A preamplifier along with a data acquisition system (DAQ) and oscilloscope are used to readout the ion signal. A blue LED external to the vacuum chamber is shone onto the windows to reduce unwanted time-varying perturbations in the experiment. In order to characterize the sensitivity, we place plates external to the vacuum system to produce electric fields at the excitation region. This induces a DC Stark shift resulting in a change in the measured ion signal.  

\subsection{Atomic beam source}

\begin{figure}
    \centering
    \includegraphics[width=0.45\textwidth]{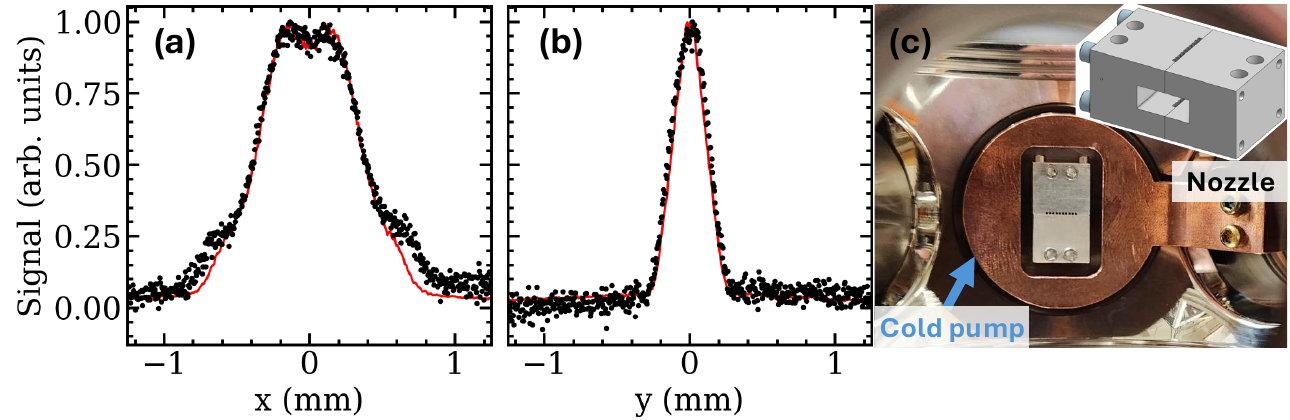}
    \caption{Spatial characteristics of the atomic beam at the location of the laser excitation region parallel (a) and perpendicular (b) to the laser propagation directions. Red curves are simulated spatial profiles while black markers are experimental values. (c) Image of the collimating nozzle and surrounding cold pump.}
    \label{fig:spatial}
\end{figure}

We have a linear array of collimated atomic beams, which together create a quasi-2D sheet of vapor. This is produced using a cascaded collimator nozzle that is similar to  Ref~\cite{Raman_collimator_2019}. In our system, a bottom reservoir is filled with natural abundance rubidium, and is heated to around 390 K. A top portion of the oven including the nozzle is heated separately to around 475 K to reduce the migration of liquid rubidium that could otherwise clog the nozzle.  The nozzle, machined from aluminum, is a two stage collimator shown in Fig.~\ref{fig:exp_setup} and Fig.~\ref{fig:spatial}(c) with 10 channels that have a total length of 9.3 mm and width and depth of ($0.5\times0.5$) mm. The channels have a spacing of 0.127 mm between them and a gap that is cut out in the center that is 3.2 mm in length. Atoms with high transverse velocity have higher probability of leaving through the gap resulting in a collimated effusive atomic beam. 

The spatial characteristics of the beam at the location of the excitation lasers, measured via fluorescence spectroscopy, is shown in Fig.~\ref{fig:spatial}(a) and (b), along with simulated spatial characteristics determined using MolFlow+ software \cite{Kersevan_MolFlow_2009}. The full width at half maximum (FWHM) of the atomic beam is approximately 8 mm  along the x-direction and 0.3 mm along the y-direction. The small size in the y-direction ensures spatial overlap of all the atoms with the laser beams. Fluorescence and absorption spectroscopy measurements show that the peak density in the beam is $\sim 10^{8}$ cm$^{-3}$. While this density is lower than that found in warm vapor cells ($\sim 10^{10}$ cm$^{-3}$ for a room temperature rubidium cell), the narrow transverse velocity spread results in reduced Doppler shifts helping improve excitation efficiency to the Rydberg state.

 Although the atomic beam is well collimated, atoms can strike a surface of the vacuum chamber, scatter, and stick to the vacuum windows. To reduce the accumulation of rubidium on the windows, two copper cold-pumps are employed. One surrounds the nozzle shown in Fig.~\ref{fig:exp_setup} and Fig.~\ref{fig:spatial}(c) while the other is placed above the beam source as shown in Fig.~\ref{fig:exp_setup}. These are cooled to around 238 K using two-stage thermo-electric coolers mounted externally to the vacuum system. The copper feedthroughs used for the cold pumps along with the nozzle are mounted on thin walled stainless steel tubing to reduce thermal contact with the rest of the vacuum chamber.

\subsection{Laser systems}

\begin{figure}
    \centering
    \includegraphics[width=0.42\textwidth]{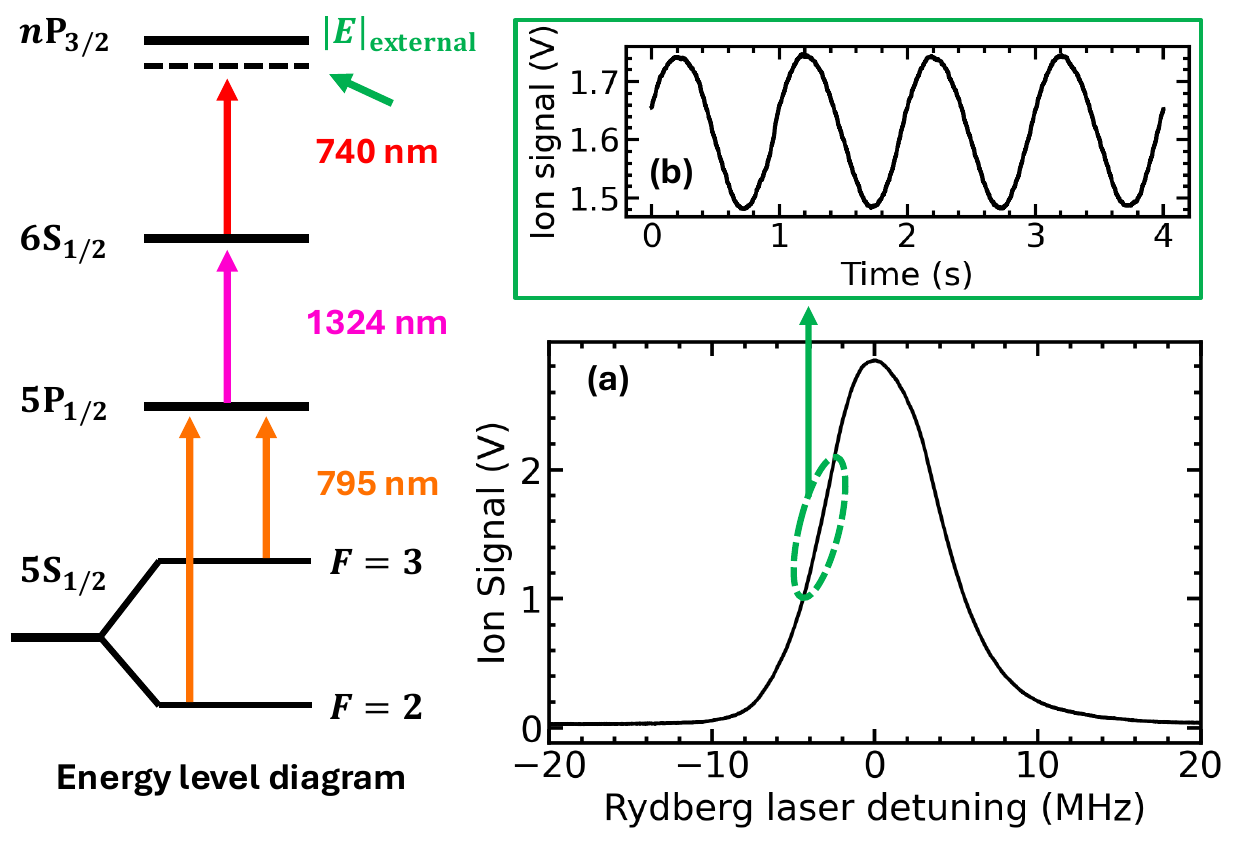}
    \caption{Energy level and laser scheme for excitation to the Rydberg state. (a) Example ion spectra for 60P$_{3/2}$. (b) Observed oscillation of the ion signal when a 1 Hz signal is applied externally to the system while the Rydberg laser is locked to the side of the ion spectra.}
    \label{fig:laser_scheme}
\end{figure}

The rubidium atoms are excited to a Rydberg state via a three photon ladder scheme shown in Fig.~\ref{fig:laser_scheme}. A ``probe'' laser at 795 nm excites the $5S_{1/2}(F=3)\rightarrow5P_{1/2}(F'=3)$ transition. A electro-optic modulator adds $3.036$ GHz frequency sidebands to the probe laser to also excite the $5S_{1/2}(F=2)\rightarrow5P_{1/2}(F'=3)$ transition, improving the total excitation efficiency to the final Rydberg state. The atoms are then excited from $5P_{1/2}(F=3)\rightarrow6S_{1/2}(F'=3)$ with a 1324 nm ``dressing" laser and finally from $6S_{1/2}(F=3)\rightarrow nP_{3/2}$ with a titanium-sapphire 740 nm ``Rydberg" laser. In this work we tune the Rydberg laser to target principle quantum numbers $n=$ 55, 60, and 101. 

To maximize the excitation efficiency the lasers are arranged in a counter-propagating orientation with the probe and dressing laser wave vector's being $\vec{k}_{\rm p} = k_{\rm p}\hat{x}$ and $\vec{k}_{\rm d}=k_{\rm d}\hat{x}$ while the Rydberg laser's wave vector is $\vec{k}_{\rm R}=-k_{\rm R}\hat{x}$ (see Fig.~\ref{fig:exp_setup}). All lasers are linearly polarized with Gaussian beam waists ($1/e^2$ value) of $1$ mm. The typical probe and dressing laser powers are about 1 mW while the Rydberg laser power is about 300 mW. The probe and dressing lasers are power stabilized using acousto-optic modulators, and frequency stabilized by locking to an ultra-stable optical cavity.

\subsection{Field ionization and detection}

Due to the excitation to a highly energetic state, the Rydberg atoms can be state selectively ionized with only modest electric fields. A simple picture in which the valence electron experiences the Coulomb potential from the nucleus summed with a 1D homogeneous ionizing potential gives reasonably accurate values for the classical ionization threshold. This occurs when the binding energy exceeds the saddle point energy requiring an electric field strength of 
\begin{equation}
    |\vec{E}|\ge \frac{\pi\epsilon_0R_y^2}{e^3n^{*4}} ,
\end{equation}
where $R_y$ is the Rydberg constant and $n^*=n-\delta_l$. Here $\delta_l$ is the quantum defect for the corresponding $n$ state. A more complete picture requires including DC Stark shifts of the state energy. Assuming an adiabatic evolution of the initial state, ionization occurs when the state energy crosses the classical ionization threshold. For $n=60$ the required field strength is $3000$ V/m while for $n=100$ it is only $350$ V/m. 

We achieve the necessary field strengths for ionization using two mesh ionization plates as shown in Fig.~\ref{fig:exp_setup}. For electrical insulation, the plates are suspended on ceramic rods, separated by 9.5 mm, and have an open area of $67\%$. The ceramic rods are mounted on stainless steel plates connected to a mesh shielding structure shown in Fig.~\ref{fig:exp_setup}. The holes shown in the side of the steel plates are used for separate experiments of laser excitation within the ionization region and are not necessary for this study. Upon passing between the plates, the Rydberg atoms are state ionized with the positive ions being deflected through the mesh and towards a CEM detector that is biased to $-1800$ V. To reduce stray fields from the ionization plates and CEM detector from reaching the laser excitation region, additional mesh shielding is placed between the laser excitation region and ionization region as shown in Fig.~\ref{fig:exp_setup}. This reduces the DC Stark shift from fringe fields of the ionization plates, and allows for a continuous operation of the system without necessitating a pulsed measurement scheme. 

To observe the ion spectra, the Rydberg laser frequency is scanned, and an example spectra for $60P_{3/2}$ is shown in Fig.~\ref{fig:laser_scheme}(a). The full width at half maximum is about 8 MHz with the dominant linewidth broadening mechanism being power broadening. Detection of external electric fields is performed by locking the Rydberg laser to the side of the ion spectra where relatively small frequency shifts $\Delta \nu$ result in a linear change in the ion signal voltage $\Delta V_{\rm ion}$. An example of this is shown in Fig.~\ref{fig:laser_scheme}(b) for a 1 Hz frequency. In this example, the ion signal from the CEM is processed with a 10 Hz low-pass filter. Frequency shifts occur due to the quadratic Stark shift of the energy level with a magnitude of \cite{Gallagher_Rydberg_1994}
\begin{equation}
    \Omega_{\rm {quad}}=-\frac{1}{2}\alpha E^2,  
\end{equation}
where $\alpha$ is the polarizability of the Rydberg state, scaling as $\alpha\propto n^7$. 

In the linear region of the ion spectra, the detectable external electric field $E_{\rm ext}$ for a quadratic Stark shift will be proportional to
\begin{equation}
    E_{\rm ext} \propto \sqrt{\frac{\Delta V_{\rm ion}}{\beta n^7}},
\end{equation}
where  $\beta$ is the slope of the ion spectra. The minimal detectable change in the ion signal voltage, $\Delta V_{\rm ion}$, will be proportional to the SNR, which in the shot noise regime is proportional to $\sqrt{N}$, where $N$ is the number of Rydberg atoms detected. As $E_{\rm ext}\propto N^{1/4}$, we can see that a 16 times increase in the number of Rydberg atoms only improves the sensitivity by a factor of two. 
\begin{figure}[t]
    \centering
    \includegraphics[width=0.48\textwidth]{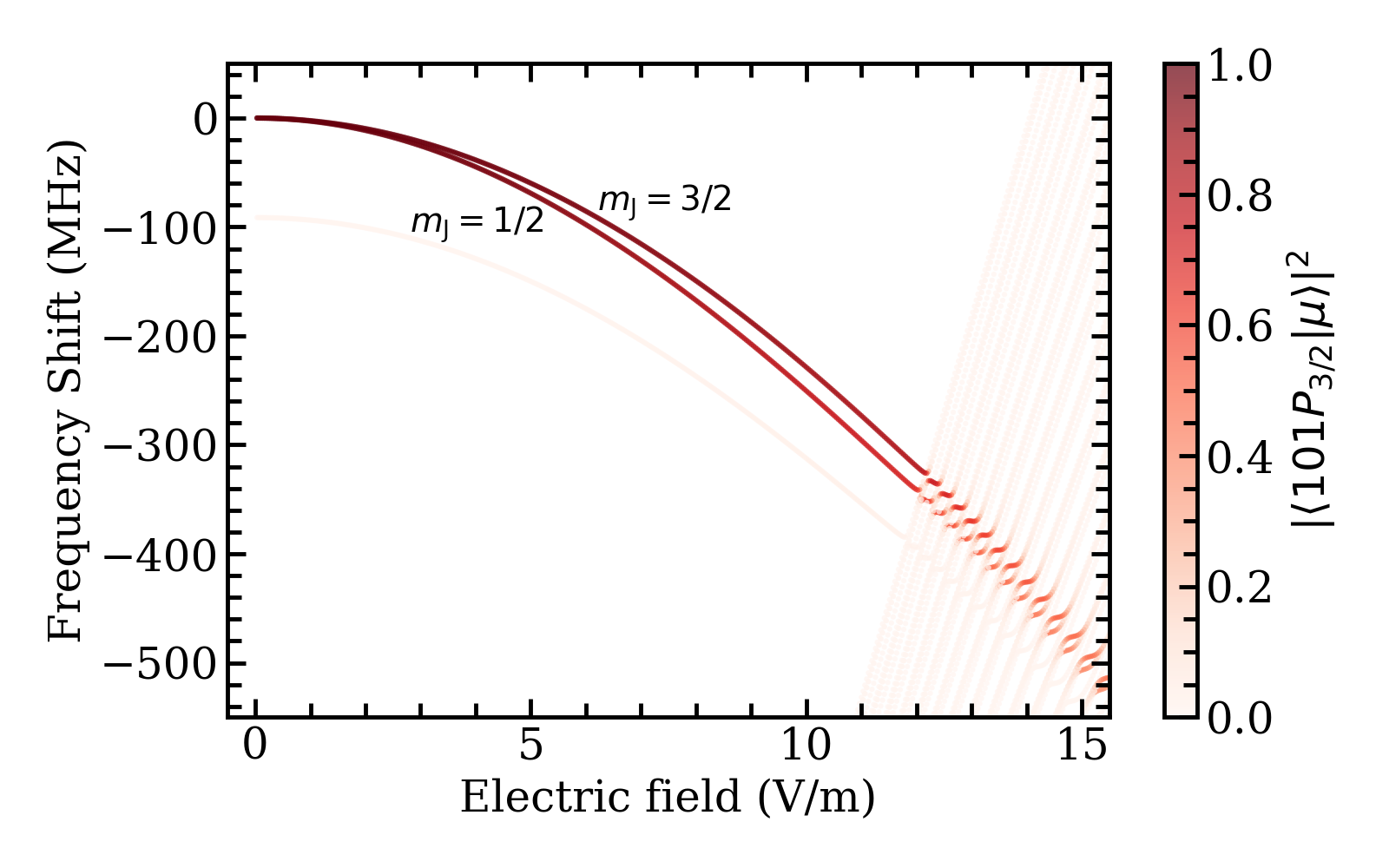}
    \caption{Theoretical Stark map of $\left|101P_{3/2},m_{\rm J}=1/2\right>$ and  $\left|101P_{3/2},m_{\rm J}=3/2\right>$ showing energy shifts and state mixing as a function of electric field amplitude. The colorbar indicates the fraction of the initial state in each atom eigenstate $\left|\mu\right>$.}
    \label{fig:state_mixing}
\end{figure}

To improve sensitivity in this quadratic Stark shift regime,  we use a bias electric field $E_{\rm b}$ where $E_{\rm b}>>E_{\rm ext}$, as has been done in prior work Ref.~\cite{Yuan_LowFreqSapphire_2020}. The total electric field amplitude at the atoms is now $E=E_{\rm b}+E_{\rm ext}$ resulting in a Stark shift of
\begin{equation}
    \Omega_{\rm {quad}}^{\rm bias}=-\frac{1}{2}\alpha \left( E_{\rm b}^2+2E_{\rm b}E_{\rm ext}+E_{\rm ext}^2\right). 
    \label{eq:bias_field_Rabi}
\end{equation}
This introduces a component of the Stark shift that is linear with respect to the external field. The minimum detectable external electric field $E_{\rm ext}^{\rm bias}$ is proportional to
\begin{equation}
    E_{\rm ext}^{\rm bias} \propto \frac{\Delta V_{\rm ion}}{E_{\rm b} \beta n^7},
    \label{eq:bias_field}
\end{equation}
where the quadratic dependence in SNR and the ion spectra slope have been removed . While it is clearly advantageous to operate with a high bias field, the scaling in Eq.~\ref{eq:bias_field} does not persist indefinitely. At sufficiently high electric field strengths, state mixing occurs leading to broadening of the ion spectra which decreases $\beta$. In Fig.~\ref{fig:state_mixing}, the theoretical Stark map for $101P_{3/2}$ is presented where state mixing can be observed beginning at around $12$ V/m due to linear Stark shifts of $n=98$ with $l=3$ to $l=97$. The Stark map is produced using the ARC-Alkali-Rydberg-Calculator python package \cite{Weatherill_ARc_2017}. Since the energy spacing between Rydberg states decreases with increasing n, the onset of state mixing and linewidth broadening will occur at lower field strengths as n increases. As such, excitation to arbitrarily high n states will also not result in an indefinite improvement to sensitivity. 

We perform sensitivity measurements at $101P_{3/2}$ and various calibration measurements at lower n states of around $n=60$. States around $n=60$ are preferable for calibrating external electric fields because splitting of the $m_{\rm J}$ states can be easily resolved, and they are relatively less susceptible to environmental noise. In fact, at n=101 we observe unintentional bias fields, including primarily from ionization plates, and also the CEM and other sources. We empirically set our controlled bias field for good sensitivity, and the unintentional fields contribute to this beneficial Stark shift, though we expect the inhomogeneity of these fields could be broadening the Rydberg ion spectra. Further improvements to the shielding and uniformity of the bias field could reduce the linewidth and improve sensitivity. 

\begin{figure}[t]
    \centering
    \includegraphics[width=0.48\textwidth]{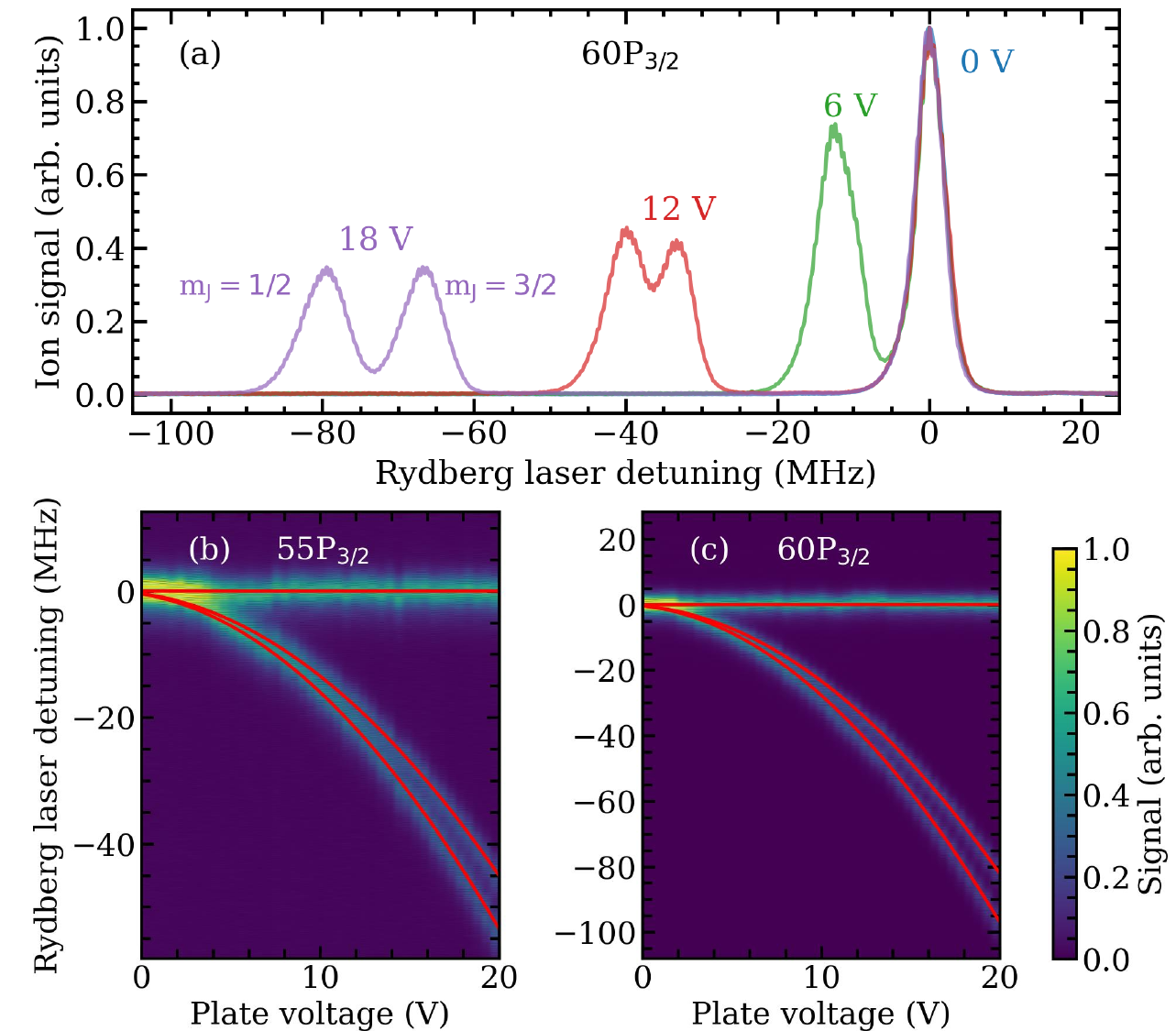}
    \caption{(a) Example DC Stark shifts and splitting of the $m_{\rm{J}}$ states for various voltages applied to the external field plates. Measured 2D Stark maps with theoretical fits (red curves) for 55P$_{3/2}$ (b) and 60P$_{3/2}$ (c). The theoretical fits are used to determine the conversion factor between the voltage applied to the external plates and electric field magnitude at the laser excitation region in the system.}
    \label{fig:field_calibration}
\end{figure} 

We empirically found that shining a blue (400 nm) LED uniformly across the windows reduced unwanted slow time-varying perturbations on the experiment. We believe the LED increases the mobility of charges on the glass, allowing shielding of slow external field noise \cite{Sedlacek2016ElectricField}. The shielding does affect the signal strength as well, and we found that for frequencies below $\approx 10$ Hz, the signal amplitude is about half of its value without the blue LED. None-the-less, the benefits of increased temporal stability outweighed the signal reduction. A more detailed study of the underlying physical dynamics of the LED-induced surface physics will be the subject of future work.

\section{Results}

\begin{figure}[t]
    \centering
    \includegraphics[width=0.35\textwidth]{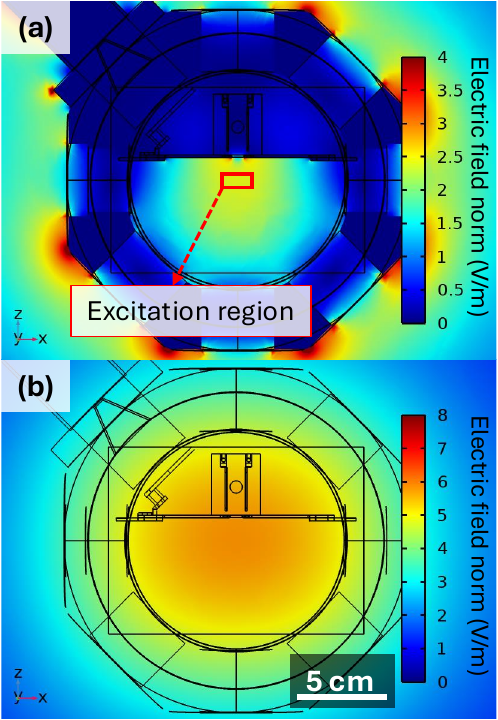}
    \caption{COMSOL simulations of the magnitude of the electric field at the center of the excitation region $(y=0)$ for 1 V applied to the external field plate. (a) Results including shielding effects from the metal chamber and internal components. (b) The metal chamber along with all internal components are treated as vacuum providing no shielding of the applied external field. The bottommost bar indicates the length scale.}
    \label{fig:comsol}
\end{figure}

In order to determine the minimum detectable electric field, it is necessary to know the electric field strength at the location of the atoms for a given voltage applied to the external field plate.  Due to shielding effects from the metal chamber and internal components this cannot be easily computed from the plate separation and applied voltage. Instead we apply a 600 Hz square wave signal to the external field plate and measure the time average ion signal as a function of Rydberg laser detuning for various plate voltages. Measuring the time averaged signal results in an effective measurement of a DC Stark shift and at sufficiently high plate voltages splitting of the $m_{\rm J}$ levels also occurs as shown in Fig.~\ref{fig:field_calibration}(a). From these results a 2D Stark map can be constructed and fit to theoretical curves where the fitting parameters provide a conversion factor from plate voltage to electric field amplitude as well as an initial Stark shift to account for residual background fields present in the chamber.

2D Stark maps and theoretical curves are given in Fig.~\ref{fig:field_calibration}(b) and (c) for $55P_{3/2}$ and $60P_{3/2}$. From the fit, the conversion parameters are such that 1 V on the external field plate corresponds to an electric field amplitude at the atoms of 
1.75(5) V/m and 1.78(5) V/m for $55P_{3/2}$ and $60P_{3/2}$
respectively. As a second means of comparison, the vacuum chamber, relevant internal components, and external field plates are modeled with the COMSOL Multiphysics software. Fig.~\ref{fig:comsol}(a) shows the simulated electric field with shielding effects from the chamber and 1 V applied to the external plate. A field strength of $\approx2.4$ V/m at the location of the atoms is predicted. We do not include possible shielding effects from surface charges on the glass windows in this model. In Fig.~\ref{fig:comsol}(b) are simulations of the field strength in which the chamber and internal metal are modeled as vacuum. This together with Fig.~\ref{fig:comsol}(a) indicates that about $55\%$ of the field strength from the plates is being shielded by the chamber.

\begin{figure}[t]
    \centering
    \includegraphics[width=0.4\textwidth]{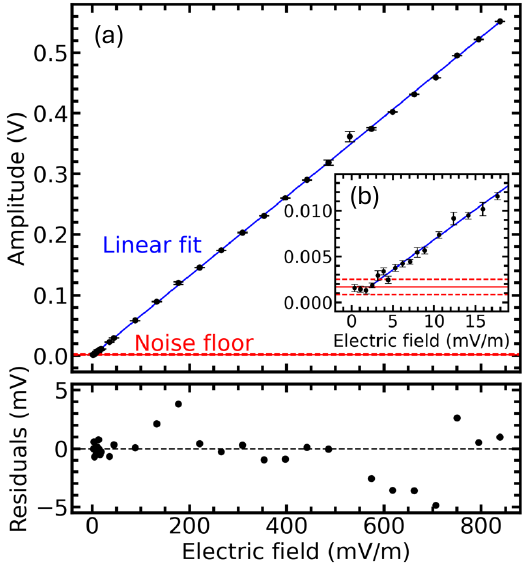}
    \caption{(a) FFT amplitude of the ion signal at 5 Hz for a 1 second measurement time versus the applied electric field. The linear portion of the data is fit to a line (blue curve), and the point where this line crosses the 5 Hz noise floor (red solid line) gives the sensitivity. The dashed red lines indicate the uncertainty in the noise floor. The inset (b) shows the FFT amplitude rolling off at lower electric field values as it approaches the noise floor.}
    \label{fig:linear_fit_ex}
\end{figure}

To determine the sensitivity of the system, we excite atoms to $101P_{\rm3/2}$ and the ion signal is recorded for 1 second. The fast Fourier transform (FFT) is then computed with a software FFT with a 1 Hz bandwidth. Over the linear response region of the ion spectra, the FFT amplitude will scale linearly with the applied electric field amplitude. Fig.~\ref{fig:linear_fit_ex}(a) shows this for a 5 Hz signal where the linear portion of the data is fit to a line. The sensitivity is given by the point where this line crosses the noise floor, or equivalently where the SNR is unity for a one second measurement. The noise floor is measured with no applied voltage to the external field plate. Fig.~\ref{fig:linear_fit_ex}(b) is zoomed in, showing the FFT amplitude rolling off as it approaches the noise floor.

\begin{figure}[t]
    \centering
    \includegraphics[width=0.48\textwidth]{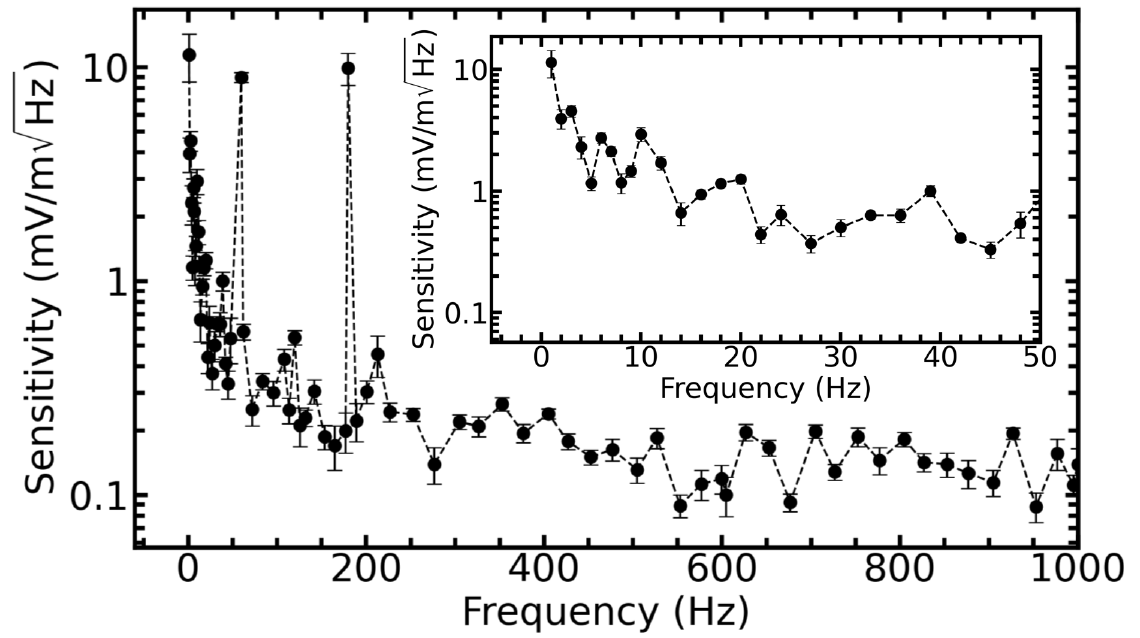}
    \caption{Measured electric field sensitivity versus electric field frequency. The inset shows the sensitivity for frequencies below 50 Hz. }
    \label{fig:min_sensitivity}
\end{figure}

The sensitivity of our sensor over a frequency range of 1 to 1200 Hz is shown in Fig.~\ref{fig:min_sensitivity}. The sensitivity is 0.14(4) mV/m$\sqrt{\rm{Hz}}$ for frequencies above 500 Hz and we see the sensitivity is  better than 1 mV/m$\sqrt{\rm {Hz}}$ for frequencies above 20 Hz. The two noticeable peaks at 60 Hz and 180 Hz are from environmental electronic noise. We note that the sensitivity in Fig.~\ref{fig:min_sensitivity} is the field strength measured at the location of the atoms and does not correct for the shielding effects from the chamber. If these chamber shielding effects were not present the sensitivity to external fields would be better by approximately a factor of two, based on our COMSOL modeling. Further, for frequencies below approximately 10 Hz, an additional factor of two improvement in sensitivity would be achieved if shielding effects from the blue LED were not present.

\begin{figure}[t]
    \centering
    \includegraphics[width=0.45\textwidth]{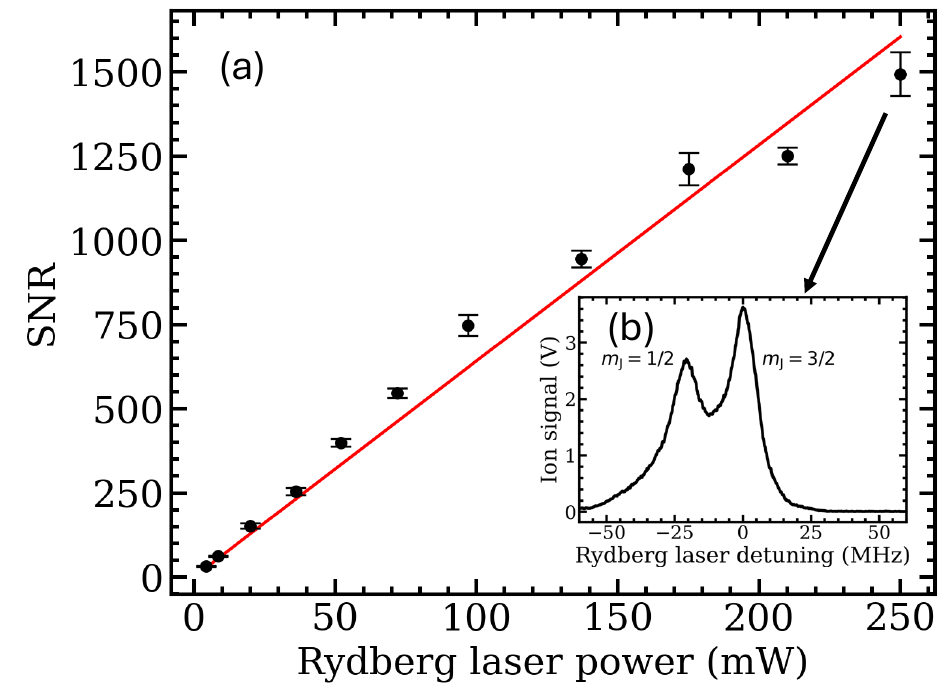}
    \caption{(a) Single shot SNR of the $101P_{3/2}, m_{\rm J}=3/2$ Rydberg ion signal versus Rydberg laser power. Data are black points fit to a red line. Error bars represent the standard deviation of the SNR from 10 shots. (b) Example single shot ion spectra for $101P_{3/2}$ with 250 mW of Rydberg laser power.}
    \label{fig:SNR_vs_power}
\end{figure}

With ionization detection we are able to achieve high single shot SNR. In Fig.~\ref{fig:SNR_vs_power}, the single shot SNR versus Rydberg laser power is presented along with an example single shot spectra for 250 mW of Rydberg laser power. This spectra showing an SNR of 1500 is obtained with a Rydberg laser frequency scan rate of 20 Hz and is only processed with our preamplifer with a low-pass 10 kHz filter. As the SNR versus Rydberg laser power follows a linear trend, we expect that with 1 W of Rydberg laser power the sensitivity would be better by a factor of 3, or around 0.05 mV/m$\sqrt{\rm{Hz}}$. It is important to note that this assumes that the sensitivity is not limited first by environmental noise. We can observe environmental noise when there is human activity close to the apparatus, and the best sensitivity is obtained during periods when there is minimal to no activity in adjacent lab spaces. 

\begin{figure}[t]
    \centering
    \includegraphics[width=0.48\textwidth]{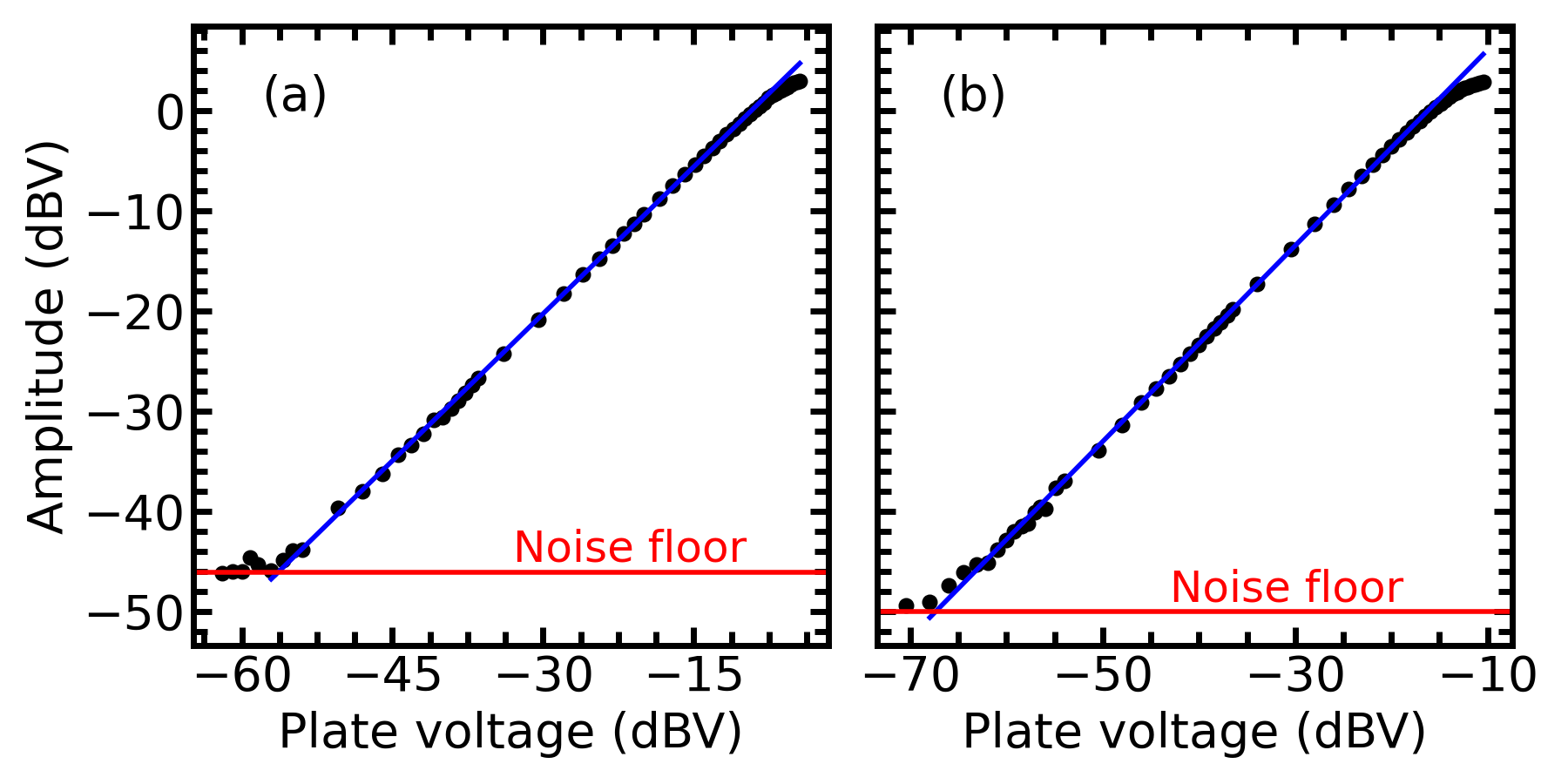}
    \caption{FFT amplitude in 1 second integration time for 20 Hz (a) and 108 Hz (b) as a function of voltage applied to the external plate. Data (black points) are fit to a line (blue) to determine the linear dynamic range, which is over 50 dB for both frequencies.}
    \label{fig:linear_dyn_range}
\end{figure}

The linear dynamic range of the system is determined by measuring the FFT amplitude for 1 second integration time and observing the point in which the linear fit exceeds the experimentally measured amplitude by 1 dB. Fig.~\ref{fig:linear_dyn_range} shows the dynamic range for a 20 Hz and 108 Hz signal applied to the external field plate. The linear dynamic range is over 50 dB for both frequencies.

\section{Conclusion}

We have presented an approach for performing low frequency electric field sensing using ionization detection on a collimated beam of Rydberg atoms. This reduces challenges associated with electric field screening effects that are present in warm alkali-metal vapor cells. With a sensitivity of $0.14(4)$ mV/m$\sqrt{\rm{Hz}}$ above 500 Hz, and ability to measure down to single Hertz frequencies, the use of a Rydberg beam and ionization detection offers a promising alternative to typical Rydberg electrometry schemes using warm vapor cells.

There are avenues for further improving the sensitivity of the current system. While we have restricted ourselves to a ``warm" atomic beam, a compact laser cooled beam source such as one from Ref.~\cite{Raman_MinitureLaserCooledBeam_2023}, would further reduce the transverse velocity spread, improving the excitation efficiency to Rydberg state while also reducing Doppler broadening. A simple comparison can be performed by numerically solving the Lindblad master equation for the laser excitation scheme discussed above to determine the steady state Rydberg population. Using the RydIQule python library \cite{Miller_2024_rydiqule, rydiqule_webpage}, we find that for an atomic beam with a transverse temperature of 10 mK, the excitation efficiency for comparable laser powers is $15$ times higher, and with a linewidth that is $3$ times narrower than our warm atomic beam. This would result in an order-of-magnitude improvement in the sensitivity, assuming the dominant linewidth broadening mechanism is not from electric field gradients. We note that at sufficiently high Rydberg densities, Rydberg-Rydberg interactions would play an important role limiting the Rydberg excitation efficiency and lifetime, and could lead to line-broadening. 

Shielding effects from the metal vacuum chamber can be reduced by using a mostly quartz vacuum chamber which have been utilized in a number of magneto-optical trap experiments \cite{Weng_GlassMOT_2020,
Boudot_GlassMOT_2020,
Bhardwaj_GlassMOT_2023}. Effects of charge accumulation on the glass surfaces can be reduced by using sapphire or sapphire coated glass windows \cite{Papoyan_GlassConductivity_1999, Yuan_LowFreqSapphire_2020}. While the interaction volume of the laser excitation region is comparable to centimeter scale vapor cells, the physical footprint of our vacuum chamber and internal components is considerably larger. The use of a miniature glass vacuum chamber in combination with a compact beam source \cite{Raman_MicroBeamSource_2020,
Martinez_MinitureBeam_2023, Raman_MinitureLaserCooledBeam_2023} would allow a significant reduction in the overall system size. Particularly in small systems, a pulsed ionization scheme could be beneficial to allow for a miniature source while avoiding unwanted Stark shifts from ionization plates in close proximity to the laser excitation region.

\begin{acknowledgments}
We gratefully acknowledge helpful discussions and input from David H. Meyer, Daniel Heinzen, Yuan-Yu Jau, and Timothy N. Nunley. We also thank Daniel Heinzen for allow us to borrow multiple pieces of equipment used in the experiment.
\end{acknowledgments}

\section{References}
\bibliographystyle{apsrev4-2}
\bibliography{refs.bib}

@article{Raman_collimator_2019,
author = {Li, Chao and Chai, Xiao and Wei, Bochao and Yang, Jeremy and Daruwalla, Anosh and Ayazi, Farrokh and Raman, C.},
address = {London},
copyright = {The Author(s) 2019},
issn = {2041-1723},
journal = {Nature communications},
number = {1},
pages = {1831-8},
publisher = {Nature Publishing Group UK},
title = {Cascaded collimator for atomic beams traveling in planar silicon devices},
volume = {10},
year = {2019},
}

@article{Biswas_geophysics_2016,
author = {Sanfui, Minu and Haldar, D. K. and Biswas, Debasish},
address = {Dordrecht},
copyright = {Springer Science+Business Media Dordrecht 2016},
issn = {0004-640X},
journal = {Astrophysics and space science},
number = {10},
pages = {1-26},
publisher = {Springer Netherlands},
title = {Studies on different geophysical and extra-terrestrial events within the Earth-ionosphere cavity in terms of ULF/ELF/VLF radio waves},
volume = {361},
year = {2016},
}

@article{Colin_geophysics_2016,
title = {ELF Electromagnetic Waves from Lightning: The Schumann Resonances},
author = {Price, Colin},
address = {BASEL},
copyright = {Copyright 2017 Elsevier B.V., All rights reserved.},
issn = {2073-4433},
journal = {Atmosphere},
number = {9},
pages = {116},
publisher = {Mdpi},
volume = {7},
year = {2016},
}

@incollection{Ali_submarine_2020,
author = {Ali, Maaruf and Miraz, Mahdi H. and Ali, Maaruf and Soomro, Safeeullah and Ware, Andrew and Miraz, Mahdi H and Excell, Peter S},
address = {Switzerland},
booktitle = {Emerging Technologies in Computing},
copyright = {ICST Institute for Computer Sciences, Social Informatics and Telecommunications Engineering 2020},
isbn = {9783030600358},
issn = {1867-8211},
pages = {86-97},
publisher = {Springer International Publishing AG},
series = {Lecture Notes of the Institute for Computer Sciences, Social Informatics and Telecommunications Engineering},
title = {A Review of Underwater Acoustic, Electromagnetic and Optical Communications},
volume = {332},
year = {2020},
}

@article{Yuan_LowFreqSapphire_2020,
  title = {Vapor-Cell-Based Atomic Electrometry for Detection Frequencies below 1 kHz},
  author = {Jau, Yuan-Yu and Carter, Tony},
  journal = {Phys. Rev. Appl.},
  volume = {13},
  issue = {5},
  pages = {054034},
  numpages = {11},
  year = {2020},
  month = {May},
  publisher = {American Physical Society},
  doi = {10.1103/PhysRevApplied.13.054034},
  url = {https://link.aps.org/doi/10.1103/PhysRevApplied.13.054034}
}

@article{Kitching_internalPlates_2022,
author = {Teale, C. and Sherman, J. and Kitching, J.},
address = {United States},
copyright = {Copyright 2022 Elsevier B.V., All rights reserved.},
issn = {2639-0213},
journal = {AVS quantum science},
number = {2},
title = {Degenerate two-photon Rydberg atom voltage reference},
volume = {4},
year = {2022},
}

@article{Suotang_internalPlates_2023,
author = {Li, Ling and Jiao, Yuechun and Hu, Jinlian and Li, Huaqiang and Shi, Meng and Zhao, Jianming and Jia, Suotang},
address = {WASHINGTON},
copyright = {Copyright 2023 Elsevier B.V., All rights reserved.},
issn = {1094-4087},
journal = {Optics express},
number = {18},
pages = {29228-29234},
publisher = {Optica Publishing Group},
title = {Super low-frequency electric field measurement based on Rydberg atoms},
volume = {31},
year = {2023},
}

@article{Meyer_assessment_2020,
author = {Meyer, David H and Castillo, Zachary A and Cox, Kevin C and Kunz, Paul D},
address = {BRISTOL},
copyright = {2020 IOP Publishing Ltd. All rights, including for text and data mining, AI training, and similar technologies, are reserved.},
issn = {0953-4075},
journal = {Journal of physics. B, Atomic, molecular, and optical physics},
number = {3},
pages = {34001-},
publisher = {IOP Publishing},
title = {Assessment of Rydberg atoms for wideband electric field sensing},
volume = {53},
year = {2020},
}

@article{Cox_MultibandDetection_2023,
  title = {Simultaneous Multiband Demodulation Using a Rydberg Atomic Sensor},
  author = {Meyer, David H. and Hill, Joshua C. and Kunz, Paul D. and Cox, Kevin C.},
  journal = {Phys. Rev. Appl.},
  volume = {19},
  issue = {1},
  pages = {014025},
  numpages = {9},
  year = {2023},
  month = {Jan},
  publisher = {American Physical Society},
  doi = {10.1103/PhysRevApplied.19.014025},
  url = {https://link.aps.org/doi/10.1103/PhysRevApplied.19.014025}
}

@article{Meyer_Polarimeter_2024,
  title = {Complete three-dimensional vector polarimetry with a Rydberg-atom rf electrometer},
  author = {Elgee, Peter K. and Cox, Kevin C. and Hill, Joshua C. and Kunz, Paul D. and Meyer, David H.},
  journal = {Phys. Rev. Appl.},
  volume = {22},
  issue = {6},
  pages = {064012},
  numpages = {11},
  year = {2024},
  month = {Dec},
  publisher = {American Physical Society},
  doi = {10.1103/PhysRevApplied.22.064012},
  url = {https://link.aps.org/doi/10.1103/PhysRevApplied.22.064012}
}

@article{Weatherill_THzImaging_2020,
  title = {Full-Field Terahertz Imaging at Kilohertz Frame Rates Using Atomic Vapor},
  author = {Downes, Lucy A. and MacKellar, Andrew R. and Whiting, Daniel J. and Bourgenot, Cyril and Adams, Charles S. and Weatherill, Kevin J.},
  journal = {Phys. Rev. X},
  volume = {10},
  issue = {1},
  pages = {011027},
  numpages = {7},
  year = {2020},
  month = {Feb},
  publisher = {American Physical Society},
  doi = {10.1103/PhysRevX.10.011027},
  url = {https://link.aps.org/doi/10.1103/PhysRevX.10.011027}
}

@article{Weatherill_THzImaging_2017,
author = {Wade, C. G. and Šibalić, N. and de Melo, N. R. and Kondo, J. M. and Adams, C. S. and Weatherill, K. J.},
address = {London},
copyright = {Springer Nature Limited 2016},
issn = {1749-4885},
journal = {Nature photonics},
number = {1},
pages = {40-43},
publisher = {Nature Publishing Group UK},
title = {Real-time near-field terahertz imaging with atomic optical fluorescence},
volume = {11},
year = {2017},
}

@article{Liang_SQL_2024,
author = {Tu, Hai Tao and Liao, Kai Yu and Wang, Hong Lei and Zhu, Yi Fei and Qiu, Si Yuan and Jiang, Hao and Huang, Wei and Bian, Wu and Yan, Hui and Zhu, Shi Liang},
address = {WASHINGTON},
issn = {2375-2548},
journal = {Science advances},
number = {51},
pages = {eads0683-},
publisher = {Amer Assoc Advancement Science},
title = {Approaching the standard quantum limit of a Rydberg-atom microwave electrometer},
volume = {10},
year = {2024},
}

@article{Adams_2007_shielding,
  title = {Coherent Optical Detection of Highly Excited Rydberg States Using Electromagnetically Induced Transparency},
  author = {Mohapatra, A. K. and Jackson, T. R. and Adams, C. S.},
  journal = {Phys. Rev. Lett.},
  volume = {98},
  issue = {11},
  pages = {113003},
  numpages = {4},
  year = {2007},
  month = {Mar},
  publisher = {American Physical Society},
  doi = {10.1103/PhysRevLett.98.113003},
  url = {https://link.aps.org/doi/10.1103/PhysRevLett.98.113003}
}

@article{Merkt_StarkShift_Cali_1999,
  title = {Using High Rydberg States as Electric Field Sensors},
  author = {Osterwalder, A. and Merkt, F.},
  journal = {Phys. Rev. Lett.},
  volume = {82},
  issue = {9},
  pages = {1831--1834},
  numpages = {0},
  year = {1999},
  month = {Mar},
  publisher = {American Physical Society},
  doi = {10.1103/PhysRevLett.82.1831},
  url = {https://link.aps.org/doi/10.1103/PhysRevLett.82.1831}
}

@article{Kersevan_MolFlow_2009,
author = {Kersevan, R. and Pons, J.-L.},
copyright = {American Vacuum Society},
issn = {0734-2101},
journal = {Journal of vacuum science and technology. A, Vacuum, surfaces, and films},
language = {eng},
number = {4},
pages = {1017-1023},
publisher = {American Vacuum Society},
title = {Introduction to MOLFLOW+: New graphical processing unit-based Monte Carlo code for simulating molecular flows and for calculating angular coefficients in the compute unified device architecture environment},
volume = {27},
year = {2009},
}

@article{Weatherill_ARc_2017,
title = {ARC: An open-source library for calculating properties of alkali Rydberg atoms},
journal = {Computer Physics Communications},
volume = {220},
pages = {319-331},
year = {2017},
issn = {0010-4655},
doi = {https://doi.org/10.1016/j.cpc.2017.06.015},
url = {https://www.sciencedirect.com/science/article/pii/S0010465517301972},
author = {N. Šibalić and J.D. Pritchard and C.S. Adams and K.J. Weatherill},
}

@article{Raithel_RF_2016,
author = {Miller, S A and Anderson, D A and Raithel, G},
address = {BRISTOL},
copyright = {2016 IOP Publishing Ltd and Deutsche Physikalische Gesellschaft},
issn = {1367-2630},
journal = {New journal of physics},
number = {5},
pages = {53017-},
publisher = {IOP Publishing},
title = {Radio-frequency-modulated Rydberg states in a vapor cell},
volume = {18},
year = {2016},
}

@article{Adams_KHz_2008,
author = {Mohapatra, Ashok K. and Bason, Mark G. and Butscher, Björn and Weatherill, Kevin J. and Adams, Charles S.},
address = {London},
copyright = {Springer Nature Limited 2008},
issn = {1745-2473},
journal = {Nature physics},
number = {11},
pages = {890-894},
publisher = {Nature Publishing Group UK},
title = {A giant electro-optic effect using polarizable dark states},
volume = {4},
year = {2008},
}

@article{Shaffer_MicrowaveElectrometry_2013,
  title = {Atom-Based Vector Microwave Electrometry Using Rubidium Rydberg Atoms in a Vapor Cell},
  author = {Sedlacek, J. A. and Schwettmann, A. and K\"ubler, H. and Shaffer, J. P.},
  journal = {Phys. Rev. Lett.},
  volume = {111},
  issue = {6},
  pages = {063001},
  numpages = {5},
  year = {2013},
  month = {Aug},
  publisher = {American Physical Society},
  doi = {10.1103/PhysRevLett.111.063001},
  url = {https://link.aps.org/doi/10.1103/PhysRevLett.111.063001}
}

@article{Shaffer_RF_sensing_calibration_2015,
author = {Fan, Haoquan and Kumar, Santosh and Sedlacek, Jonathon and Kübler, Harald and Karimkashi, Shaya and Shaffer, James P},
address = {BRISTOL},
copyright = {2015 IOP Publishing Ltd},
issn = {0953-4075},
journal = {Journal of physics. B, Atomic, molecular, and optical physics},
number = {20},
pages = {202001-202016},
publisher = {IOP Publishing},
title = {Atom based RF electric field sensing},
volume = {48},
year = {2015},
}

@ARTICLE{Bonnie_RydbergSensing_2021,
  author={Fancher, Charles T. and Scherer, David R. and John, Marc C. St. and Marlow, Bonnie L. Schmittberger},
  journal={IEEE Transactions on Quantum Engineering}, 
  title={Rydberg Atom Electric Field Sensors for Communications and Sensing}, 
  year={2021},
  volume={2},
  number={},
  pages={1-13},
  doi={10.1109/TQE.2021.3065227}}

@article{Holloway_Calibration_2024,
author = {Schlossberger, Noah and Prajapati, Nikunjkumar and Berweger, Samuel and Rotunno, Andrew P. and Artusio-Glimpse, Alexandra B. and Simons, Matthew T. and Sheikh, Abrar A. and Norrgard, Eric B. and Eckel, Stephen P. and Holloway, Christopher L.},
address = {London},
copyright = {This is a U.S. Government work and not under copyright protection in the US; foreign copyright protection may apply 2024. corrected publication 2024},
issn = {2522-5820},
journal = {Nature reviews physics},
number = {10},
pages = {606-620},
publisher = {Nature Publishing Group UK},
title = {Rydberg states of alkali atoms in atomic vapour as SI-traceable field probes and communications receivers},
volume = {6},
year = {2024},
}

@article{Arimondo_RbMotShielding_2011,
author = {Viteau, M. and Radogostowicz, J. and Bason, M. G. and Malossi, N. and Ciampini, D. and Morsch, O. and Arimondo, E.},
address = {WASHINGTON},
copyright = {Copyright 2014 Elsevier B.V., All rights reserved.},
issn = {1094-4087},
journal = {Optics express},
number = {7},
pages = {6007-6019},
publisher = {Optical Soc Amer},
title = {Rydberg spectroscopy of a Rb MOT in the presence of applied or ion created electric fields},
volume = {19},
year = {2011},
}

@article{Gallagher_RydbergInteraction_1981,
  title = {Resonant Rydberg-Atom-Rydberg-Atom Collisions},
  author = {Safinya, K. A. and Delpech, J. F. and Gounand, F. and Sandner, W. and Gallagher, T. F.},
  journal = {Phys. Rev. Lett.},
  volume = {47},
  issue = {6},
  pages = {405--408},
  numpages = {0},
  year = {1981},
  month = {Aug},
  publisher = {American Physical Society},
  doi = {10.1103/PhysRevLett.47.405},
  url = {https://link.aps.org/doi/10.1103/PhysRevLett.47.405}
}

@article{Gallagher_RydbergInteraction_1998,
  title = {Resonant Dipole-Dipole Energy Transfer in a Nearly Frozen Rydberg Gas},
  author = {Anderson, W. R. and Veale, J. R. and Gallagher, T. F.},
  journal = {Phys. Rev. Lett.},
  volume = {80},
  issue = {2},
  pages = {249--252},
  numpages = {0},
  year = {1998},
  month = {Jan},
  publisher = {American Physical Society},
  doi = {10.1103/PhysRevLett.80.249},
  url = {https://link.aps.org/doi/10.1103/PhysRevLett.80.249}
}

@article{Vanhaecke_RydbergInteractions_2010,
  title = {Landau-Zener Transitions in Frozen Pairs of Rydberg Atoms},
  author = {Saquet, Nicolas and Cournol, Anne and Beugnon, J\'er\^ome and Robert, Jacques and Pillet, Pierre and Vanhaecke, Nicolas},
  journal = {Phys. Rev. Lett.},
  volume = {104},
  issue = {13},
  pages = {133003},
  numpages = {4},
  year = {2010},
  month = {Apr},
  publisher = {American Physical Society},
  doi = {10.1103/PhysRevLett.104.133003},
  url = {https://link.aps.org/doi/10.1103/PhysRevLett.104.133003}
}

@article{Tannian_StarkShift_1999,
author = {Tannian, B. E. and Stokely, C. L. and Dunning, F. B. and Reinhold, C. O. and Burgdörfer, J.},
address = {BRISTOL},
copyright = {Copyright 2004 Elsevier Science B.V., Amsterdam. All rights reserved.},
issn = {0953-4075},
journal = {Journal of physics. B, Atomic, molecular, and optical physics},
number = {18},
pages = {L517-L524},
publisher = {Iop Publishing Ltd},
title = {Manipulation of atomic l-state distributions using pulsed electric fields},
volume = {32},
year = {1999},
}

@article{Grimmel_StarkShift_2017,
  title = {Ionization spectra of highly Stark-shifted rubidium Rydberg states},
  author = {Grimmel, Jens and Stecker, Markus and Kaiser, Manuel and Karlewski, Florian and Torralbo-Campo, Lara and G\"unther, Andreas and Fort\'agh, J\'ozsef},
  journal = {Phys. Rev. A},
  volume = {96},
  issue = {1},
  pages = {013427},
  numpages = {7},
  year = {2017},
  month = {Jul},
  publisher = {American Physical Society},
  doi = {10.1103/PhysRevA.96.013427},
  url = {https://link.aps.org/doi/10.1103/PhysRevA.96.013427}
}

@article{Wallraff_Cryogenic_2015,
  title = {Imaging electric fields in the vicinity of cryogenic surfaces using Rydberg atoms},
  author = {Thiele, T. and Deiglmayr, J. and Stammeier, M. and Agner, J.-A. and Schmutz, H. and Merkt, F. and Wallraff, A.},
  journal = {Phys. Rev. A},
  volume = {92},
  issue = {6},
  pages = {063425},
  numpages = {6},
  year = {2015},
  month = {Dec},
  publisher = {American Physical Society},
  doi = {10.1103/PhysRevA.92.063425},
  url = {https://link.aps.org/doi/10.1103/PhysRevA.92.063425}
}

@article{Papoyan_GlassConductivity_1999,
author = {Bouchiat, M.A. and Guéna, J. and Jacquier, Ph and Lintz, M. and Papoyan, A.V.},
address = {NEW YORK},
copyright = {Copyright 2007 Elsevier B.V., All rights reserved.},
issn = {0946-2171},
journal = {Applied physics. B, Lasers and optics},
number = {6},
pages = {1109-1116},
publisher = {Springer Nature},
title = {Electrical conductivity of glass and sapphire cells exposed to dry cesium vapor},
volume = {68},
year = {1999},
}

@article{Bhardwaj_GlassMOT_2023,
author = {Bhardwaj, Kavish and Sarkar, Sourabh and Ram, S. P. and Tiwari, V. B. and Mishra, S. R.},
address = {MELVILLE},
copyright = {Author(s)},
issn = {2158-3226},
journal = {AIP advances},
number = {1},
pages = {015108-015108-5},
publisher = {AIP Publishing},
title = {A method for loading magneto-optical trap in an ultrahigh vacuum environment},
volume = {13},
year = {2023},
}

@article{Boudot_GlassMOT_2020,
author = {Boudot, Rodolphe and McGilligan, James P. and Moore, Kaitlin R. and Maurice, Vincent and Martinez, Gabriela D. and Hansen, Azure and de Clercq, Emeric and Kitching, John},
address = {London},
copyright = {The Author(s) 2020},
issn = {2045-2322},
journal = {Scientific reports},
number = {1},
pages = {16590-},
publisher = {Nature Publishing Group UK},
title = {Enhanced observation time of magneto-optical traps using micro-machined non-evaporable getter pumps},
volume = {10},
year = {2020},
}

@article{Weng_GlassMOT_2020,
author = {Weng, Kanxing and Wu, Bin and Lin, Jiahong and Zhou, Yin and Cheng, Bing and Lin, Qiang},
address = {WASHINGTON},
copyright = {Copyright 2020 Elsevier B.V., All rights reserved.},
issn = {0740-3224},
journal = {Journal of the Optical Society of America. B, Optical physics},
number = {6},
pages = {1637-1642},
publisher = {Optical Soc Amer},
title = {Compact magneto-optical trap with a quartz vacuum chamber for miniature gravimeters},
volume = {37},
year = {2020},
}

@article{Raman_MinitureLaserCooledBeam_2023,
  title = {Stimulated Laser Cooling in a Compact Geometry Using Microfabricated Atomic Beam Collimators},
  author = {Li, Chao and Chai, Xiao and Zhuo, Linzhao and Wei, Bochao and Lotfi, Ardalan and Ayazi, Farrokh and Raman, Chandra},
  journal = {Phys. Rev. Appl.},
  volume = {20},
  issue = {3},
  pages = {034042},
  numpages = {13},
  year = {2023},
  month = {Sep},
  publisher = {American Physical Society},
  doi = {10.1103/PhysRevApplied.20.034042},
  url = {https://link.aps.org/doi/10.1103/PhysRevApplied.20.034042}
}

@article{Martinez_MinitureBeam_2023,
author = {Martinez, Gabriela D. and Li, Chao and Staron, Alexander and Kitching, John and Raman, Chandra and McGehee, William R.},
address = {London},
copyright = {This is a U.S. Government work and not under copyright protection in the US; foreign copyright protection may apply 2023},
issn = {2041-1723},
journal = {Nature communications},
number = {1},
pages = {3501-7},
publisher = {Nature Publishing Group UK},
title = {A chip-scale atomic beam clock},
volume = {14},
year = {2023},
}

@article{Raman_MicroBeamSource_2020,
  title = {Robust characterization of microfabricated atomic beams on a six-month time scale},
  author = {Li, Chao and Wei, Bochao and Chai, Xiao and Yang, Jeremy and Daruwalla, Anosh and Ayazi, Farrokh and Raman, C.},
  journal = {Phys. Rev. Res.},
  volume = {2},
  issue = {2},
  pages = {023239},
  numpages = {14},
  year = {2020},
  month = {May},
  publisher = {American Physical Society},
  doi = {10.1103/PhysRevResearch.2.023239},
  url = {https://link.aps.org/doi/10.1103/PhysRevResearch.2.023239}
}

@book{Gallagher_Rydberg_1994,
author = {Gallagher, Thomas F.},
address = {Cambridge},
booktitle = {Rydberg atoms},
isbn = {0-511-52453-6},
publisher = {Cambridge University Press},
series = {Cambridge monographs on atomic, molecular, and chemical physics ; 3},
title = {Rydberg atoms / Thomas F. Gallagher.},
year = {1994},
}

@article{Miller_2024_rydiqule,
title = {RydIQule: A graph-based paradigm for modeling Rydberg and atomic sensors},
journal = {Computer Physics Communications},
volume = {294},
pages = {108952},
year = {2024},
issn = {0010-4655},
doi = {https://doi.org/10.1016/j.cpc.2023.108952},
url = {https://www.sciencedirect.com/science/article/pii/S0010465523002977},
author = {Benjamin N. Miller and David H. Meyer and Teemu Virtanen and Christopher M. O'Brien and Kevin C. Cox},
}

@misc{rydiqule_webpage,
  howpublished = {\url{https://rydiqule.readthedocs.io/en/stable/index.html}},
}

@article{Kleppner_SFI_1975,
  title = {Stark Ionization of High-Lying States of Sodium},
  author = {Ducas, Theodore W. and Littman, Michael G. and Freeman, Richard R. and Kleppner, Daniel},
  journal = {Phys. Rev. Lett.},
  volume = {35},
  issue = {6},
  pages = {366--369},
  numpages = {0},
  year = {1975},
  month = {Aug},
  publisher = {American Physical Society},
  doi = {10.1103/PhysRevLett.35.366},
  url = {https://link.aps.org/doi/10.1103/PhysRevLett.35.366}
}

@article{Edelstein_SFI_1977,
  title = {Field ionization of highly excited states of sodium},
  author = {Gallagher, T. F. and Humphrey, L. M. and Cooke, W. E. and Hill, R. M. and Edelstein, S. A.},
  journal = {Phys. Rev. A},
  volume = {16},
  issue = {3},
  pages = {1098--1108},
  numpages = {0},
  year = {1977},
  month = {Sep},
  publisher = {American Physical Society},
  doi = {10.1103/PhysRevA.16.1098},
  url = {https://link.aps.org/doi/10.1103/PhysRevA.16.1098}
}

@article{Cassidy_SFI_2018,
  title = {State-selective electric-field ionization of Rydberg positronium},
  author = {Alonso, A. M. and Gurung, L. and Sukra, B. A. D. and Hogan, S. D. and Cassidy, D. B.},
  journal = {Phys. Rev. A},
  volume = {98},
  issue = {5},
  pages = {053417},
  numpages = {12},
  year = {2018},
  month = {Nov},
  publisher = {American Physical Society},
  doi = {10.1103/PhysRevA.98.053417},
  url = {https://link.aps.org/doi/10.1103/PhysRevA.98.053417}
}

@article{Sedlacek2016ElectricField,
  title        = {Electric Field Cancellation on Quartz by Rb Adsorbate-Induced Negative Electron Affinity},
  author       = {Sedlacek, J. A. and Kim, E. and Rittenhouse, S. T. and Weck, P. F. and Sadeghpour, H. R. and Shaffer, J. P.},
  journal      = {Physical Review Letters},
  volume       = {116},
  number       = {13},
  pages        = {133201},
  year         = {2016},
  publisher    = {American Physical Society},
  doi          = {10.1103/PhysRevLett.116.133201},
  url          = {https://doi.org/10.1103/PhysRevLett.116.133201}
}

@article{Kunz_OptimalEIT_2021,
  title = {Optimal atomic quantum sensing using electromagnetically-induced-transparency readout},
  author = {Meyer, David H. and O'Brien, Christopher and Fahey, Donald P. and Cox, Kevin C. and Kunz, Paul D.},
  journal = {Phys. Rev. A},
  volume = {104},
  issue = {4},
  pages = {043103},
  numpages = {11},
  year = {2021},
  month = {Oct},
  publisher = {American Physical Society},
  doi = {10.1103/PhysRevA.104.043103},
  url = {https://link.aps.org/doi/10.1103/PhysRevA.104.043103}
}

@article{Cappellaro_QuantumSensing_2017,
  title = {Quantum sensing},
  author = {Degen, C. L. and Reinhard, F. and Cappellaro, P.},
  journal = {Rev. Mod. Phys.},
  volume = {89},
  issue = {3},
  pages = {035002},
  numpages = {39},
  year = {2017},
  month = {Jul},
  publisher = {American Physical Society},
  doi = {10.1103/RevModPhys.89.035002},
  url = {https://link.aps.org/doi/10.1103/RevModPhys.89.035002}
}

@article{Facon_SensitiveElectrometer_2016,
  title        = {A sensitive electrometer based on a Rydberg atom in a Schrödinger-cat state},
  author       = {Facon, Adrien and Dietsche, Eva-Katharina and Grosso, Dorian and Haroche, Serge and Raimond, Jean-Michel and Brune, Michel and Gleyzes, Sébastien},
  journal      = {Nature},
  volume       = {535},
  number       = {7611},
  pages        = {262--265},
  year         = {2016},
  doi          = {10.1038/nature18327},
  url          = {https://doi.org/10.1038/nature18327}
}

@misc{Chandra_ELF_2026,
      title={Electrometry of extremely-low frequencies from kHz to sub-Hz with a Rydberg-atom sensor}, 
      author={Aveek Chandra and Narongrit Paensin and Rainer Dumke},
      year={2026},
      eprint={2603.13827},
      archivePrefix={arXiv},
      primaryClass={quant-ph},
      url={https://arxiv.org/abs/2603.13827}, 
}

@misc{Yuan_ELF_2026,
      title={Very sensitive vapor-cell quasi-DC atomic E-field sensor}, 
      author={Amy Damitz and George Burns and Yuan-Yu Jau},
      year={2026},
      eprint={2603.23751},
      archivePrefix={arXiv},
      primaryClass={physics.atom-ph},
      url={https://arxiv.org/abs/2603.23751}, 
}

@misc{Kayim_lowFreq_2026,
      title={Calibration of electric fields in low-frequency off-resonant Rydberg receivers}, 
      author={Baran Kayim and Michael A. Viray and David S. La Mantia and Daniel Richardson and James Dee and Ryan S. Westafer and Brian C. Sawyer and Robert Wyllie},
      year={2026},
      eprint={2603.10898},
      archivePrefix={arXiv},
      primaryClass={physics.atom-ph},
      url={https://arxiv.org/abs/2603.10898}, 
}

\end{document}